\documentclass[a4paper,useAMS,usenatbib]{mnras}

\usepackage{graphicx}
\usepackage{setspace}
\usepackage{natbib}
\usepackage{color}
\usepackage{amsmath,amssymb}
\usepackage{times}
\usepackage{hyperref}

\title[Clouds in Arms]{Clouds in Arms}

\author[V. A. Belokurov \& D. Erkal]{Vasily A. Belokurov$^{1,2}$\thanks{E-mail:vasily@ast.cam.ac.uk} \& Denis Erkal$^{3}$\\
  $^{1}$Institute of Astronomy, Madingley Rd, Cambridge, CB3 0HA\\
  $^{2}$Center for Computational Astrophysics, Flatiron Institute, 162 5th Avenue, New York, NY 10010, USA\\
  $^{3}$Department of Physics, University of Surrey, Guildford GU2 7XH, UK\\
}

\begin{document}

\maketitle

\label{firstpage}

\begin{abstract}
We use astrometry and broad-band photometry from Data Release 2 of the
ESA's Gaia mission to map out low surface-brightness features in the
stellar density distribution around the Large and Small Magellanic
Clouds. The LMC appears to have grown two thin and long stellar
streams in its Northern and Southern regions, highly reminiscent of
spiral arms. We use computer simulations of the Magellanic Clouds'
in-fall to demonstrate that these arms were likely pulled out of the
LMC's disc due to the combined influence of the SMC's most recent
fly-by and the tidal field of the Milky Way.

\end{abstract}


\section{Introduction}

Stellar discs are fragile and even a quick, low mass-ratio encounter
with a neighboring galaxy can cause plenty of damage. The blockbuster
by \citet{Toomre1972} provides a gallery of salient moments of such
interactions as well as a comprehensive analysis of plausible
outcomes. Let us provide a digest of their findings as to the
formation of arms and bridges between companion galaxies. First,
\citet{Toomre1972} point out that the damage is inflicted via tidal
forces, which are roughly symmetric with respect to the disc's
host. Thus, a single passage will always produce two arms (whose
relative strengths depend on the perturber's orbit) on opposite sides
of the disc.  No slow interaction is needed, relatively fast
(parabolic) orbits will also lead to arm formation. Naturally, smaller
perturbers pull out tidier, i.e. more coherent arms as the fly-by of a
massive neighbor causes a messier debris splatter. However, smaller
perturbers take more time to pull out long arms and have to come
closer to the disc compared to the massive ones. \citet{Toomre1972}
highlight repeatedly how narrow the tidally-induced arms are, but take
care to point out that this thinness is quite often the result of the
perspective, in fact most arms are ``ribbons'', not ``strings''. While
arm production can be thought of as a resonance phenomenon \citep[see
  aslo][]{Elena2010}, even highly inclined encounters produce dramatic
arms. In the latter case, arms usually twist considerably in 3D, and
while appearing face-on€ for some viewing angles are clearly pulled
out of the disc plane.

While the study of \citet{Toomre1972} is motivated by such iconic
images as that of e.g. M51, one can find several {\it very local}
examples of low mass-ratio galaxy conflicts with dramatic
consequences. Most notably, as described in \citet{Laporte2017} and
\citet{Laporte2018}, the Sgr dwarf - itself barely a twentieth of the
Milky Way's mass - has likely wrought plenty of havoc in the Galaxy's
disc. The dwarf is now held responsible for inducing a large-scale
spiral structure in the Galaxy \citep[see e.g.][]{Purcell2011},
creating a warp in the gaseous disc \citep[see e.g.][]{Gibbons2017}
and sending large-amplitude waves through the stellar disc
\citep[e.g.][]{Widrow2012,Ralph2017,Xu2015}. Most interesting are the
long thin streams of stars likely pulled out of the Galactic disc
\citep[see][]{Carl2006,Carl2011,deBoer2018,Deason2018} that do look
remarkably similar to the tidal arms described in \citet{Toomre1972}
and that can now be used for a variety of chemo-dymamical studies of
both the Milky Way and the Sgr dwarf \citep[see][]{Laporte2018}.

\begin{figure*}
  \centering
  \includegraphics[width=0.95\textwidth]{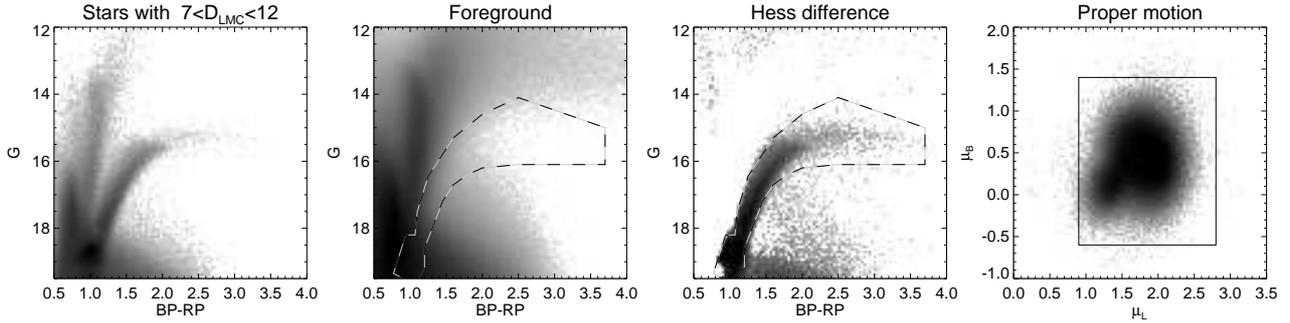}
  \caption[]{The Magellanic Clouds in Gaia DR2. Only stars with
    parallax $\varpi<0.2$ mas are used. {\it Left:} Logarithm of the
    density of the outer LMC stars in $G$ vs $G_{\rm
      BP}-G_{\rm RP}$ space (all de-reddened). {\it Middle Left:} Logarithm
    of the density of stars in the Galactic foreground near the
    Clouds. {\it Middle Right:} The difference of the CMD densities
    shown in the first two panels. Apart from some minor contamination
    at faint $G$ and red $G_{\rm BP}-G_{\rm RP}$, the strongest
    over-density is that corresponding to the LMC's red giant
    branch. The black-white dashed line shows the CMD mask used to
    select the likely LMC (and SMC) giants. {\it Right:} Logarithm of
    stellar density in proper motion space. In addition to the CMD
    selection shown in the middle panels, we select stars within
    $15^{\circ}$ ($10^{\circ}$) of the LMC's (SMC's) center and with
    $G_{\rm BP}-G_{\rm RP} > 1.3$. Black lines outline the proper
    motion selection box used to improve purity of the tracer
    population.}
   \label{fig:selection}
\end{figure*}

It so happened that the most striking example of a nearby binary
interaction was only just being discovered at the time of the writing of
\citet{Toomre1972} and hence could not be included in their
analysis. \citet{Wannier1972} and \citet{Kuilenburg1972} detected long
streams of HI in the Southern sky, and some two years later these were
shown to connect to the Magellanic Clouds by
\citet{Mathewson1974}. The Magellanic Stream (as it is known today)
has since been mapped across the sky \citep[see
  e.g.][]{Putman2003,Nidever2008,Nidever2010} and is today
unambiguously demonstrated to have originated in the interaction
between the Large and the Small Clouds \citep[][]{Besla2007,
  Besla2010, Diaz2011, Diaz2012}. While the stellar counterpart to the
MS is yet to be discovered, the last two years have seen a marked
increase in the number of reported detections of low
surface-brightness stellar sub-structure in the vicinity of the Clouds
\citep[see
  e.g.][]{Mackey2016,Belokurov2016,Belokurov2017,Deason2017,Pieres2017,Mackey2018,Nidever2018}. In
particular, \citet{Mackey2016,Mackey2018} concentrate on the
perturbations in and around the LMC's stellar disc. They uncover a
wealth of sub-structure, some of which (such as the long stream-like
feature in the North of the LMC) they tentatively attribute to the
tidal influence of the MW \citep[][]{Mackey2016}. They also detect
prominent stellar debris over-densities in the Southern parts of the
LMC and put forward two formation scenarios: one to do with the
disruption of the LMC's disc and one linked to the episodic stripping
of the SMC \citep[see also][who argued for the importance of repeated
  interactions with the SMC]{besla_etal_2016}.

\begin{figure*}
  \centering
  \includegraphics[width=0.97\textwidth]{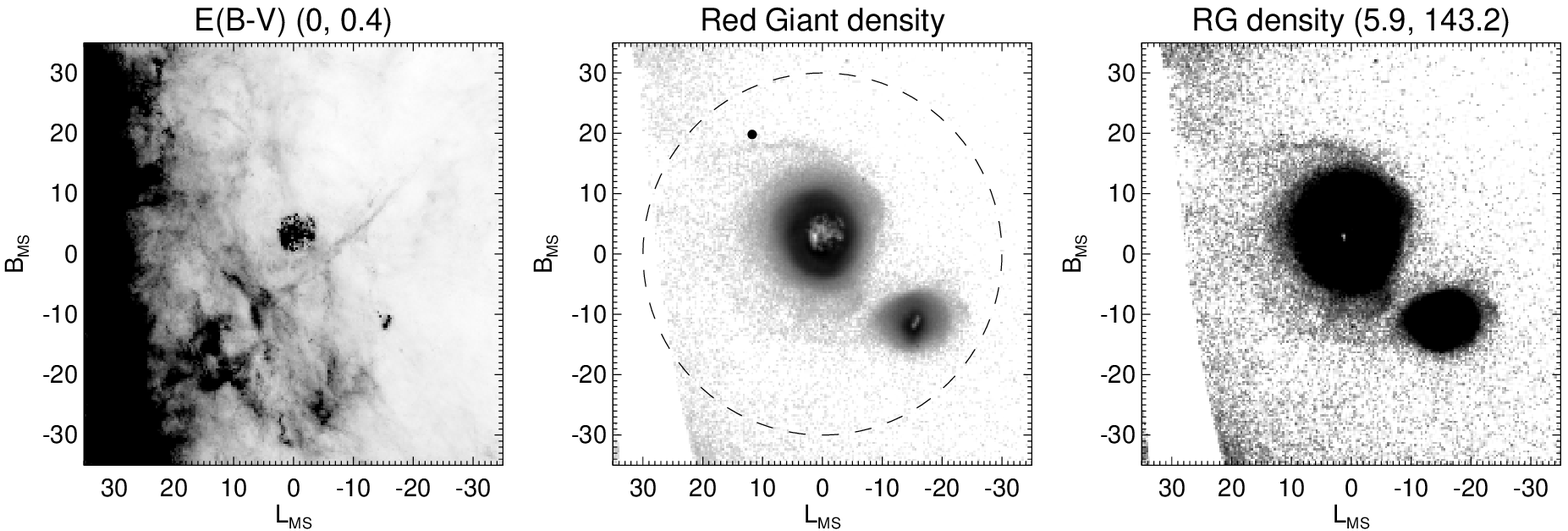}
  \vskip -0.1cm
  \caption[]{ {\it Left:} Map of the distribution of the total dust
    extinction centered on the LMC as measured by \citet{SFD} {\it
      Middle:} Density of the candidate RGB stars selected using cuts
    illustrated in Figure~\ref{fig:selection} and described in the
    main text. Filled black circle marks the location of the Carina
    dwarf spheroidal galaxy. {\it Right:} Same as Middle but saturated
    at lower density levels. Number of stars per square degree
    corresponding to the white (low density) and black (high density)
    is given in the title.}
   \label{fig:map}
\end{figure*}

In this Letter, we use a combination of Gaia's (Data Release 2, or
GDR2) photometry and astrometry to produce an uninterrupted panorama
of the Magellanic Clouds. We focus on the density fluctuations between
10 and 30 degrees away from the LMC's centre. While our maps do not
attain the same level of detail achievable using deep imaging with
instruments such as DECam, they help to fill in the gaps in the
Magellanic puzzle. Moreover, Gaia's astrometry has the
unprecedented power to remove the bulk of the intervening Milky Way's
disc population and thus extend the study of the Clouds to regions
not accessible even with the deepest imaging surveys. Specifically, we
demonstrate that two long and narrow tidal arms exist in the Northern
and Southern outskirts of the LMC's disc, most likely produced as a
result of the combined effect of the MW tides and the interaction with
the SMC during its most recent passage near the Large Cloud.

\section{Gaia DR2 view of the Magellanic Clouds}

In what follows we use the photometry and astrometry provided as part
of the Data Release 2 \citep[][]{Brown2018} of the Gaia mission
\citep[][]{Prusti2016}. We correct the $G, G_{\rm BP}$ and $G_{\rm
  RP}$ magnitudes for the effects of extinction using the first two
terms in the Equation 1 of \citet{Babusiaux2018} and the dust maps of
\citet{SFD}. Additionally, we remove the foreground dwarf stars from
our sample by culling all objects with parallax $\varpi>0.2$ mas and
exclude stars with Galactic latitudes $|b|<5^{\circ}$. We note that
this is not the first attempt to use GDR2 to study the LMC (and the
SMC): the kinematic view of the inner portions of each Cloud can be
found in \citet{Helmi2018}, while \citet{Vasiliev2018} presents the
first results of dynamical modelling of the inner LMC.

Figure ~\ref{fig:selection} shows the behavior of stars with
$\varpi<0.2$ mas in the vicinity of the Clouds in color-magnitude and
proper motion spaces. More precisely, the left panel displays the
density of stars within 12 degrees of the LMC's center in $G$ vs
$G_{\rm BP}-G_{\rm RP}$ plane (Hess diagram). Here we assumed that the
center of the dwarf is located at $\alpha, \delta =
80.89375^{\circ},-69.7561^{\circ}$. The CMD signal of the LMC can be
compared to that of the Galactic foreground shown in the second panel
of the Figure. We give the difference of the two in the third
panel. In this Hess difference plot, the LMC's Red Giant Branch (RGB)
and the Red Clump (RC) are easily discernible (their envelope is
traced by black-and-white dashed line). Note that the tip of the RGB
runs horizontally (i.e. at constant $G$) for colors redder than
$G_{\rm BP}-G_{\rm RP} \simeq 2$. While the RC is the most densely
populated CMD feature, it is also the one that suffers the highest
Galactic foreground contamination, especially at $G_{\rm BP}-G_{\rm
  RP}<0.9$ and $G>19$. Therefore, to select the likely Magellanic
stars we choose objects with $\varpi<0.2$ mas that fall within the CMD
mask (broad enough to accommodate the heliocentric distance range
across the Magellanic system) shown in panels 2 and 3 of
Figure~\ref{fig:selection} and have $G_{\rm BP}-G_{\rm RP}>0.9$ and
$G<19$. Finally, to further improve the purity of our selection we
apply proper motion cuts chosen to delineate the motion of genuine LMC
and SMC stars as shown in the fourth (rightmost) panel of the
Figure. Here, stars within 15$^{\circ}$ of the LMC and 7$^{\circ}$ of
the SMC are shown in $\mu_{\rm L}, \mu_{\rm B}$ proper motion space
aligned with the gaseous MS \citep[see][for the definition of the
  $L_{\rm MS}, B_{\rm MS}$ coordinate system]{Nidever2008}. Note that
to clarify the over-densities corresponding to the Clouds, for this
panel only, we additionally limit the stars to those with $G_{\rm
  BP}-G_{\rm RP}>1.3$.

The density of the likely Magellanic RGB candidate stars selected
using a combination of parallax, $|b|$, CMD and proper motion cuts
described above is shown in Figure~\ref{fig:map}. The same density map
is displayed twice, in the middle and right panels of the Figure,
albeit with different saturation levels to help study features across
a wide range of surface brightness values. Note that even at
astonishingly low Galactic latitudes, $|b|<10^{\circ}$, very little
disc contamination is visible thanks to the power of Gaia's
astrometry. Comparing the stellar density patterns in panels 2 and 3
with the dust distribution shown in panel 1, we conclude that the
only noticeable correlation between the two maps can be seen in the
very cores of each Cloud, where the star counts are depleted by high
values of extinction. Figure~\ref{fig:map} reveals an intricate and
spatially extended system of narrow stream-like structures emanating
from the LMC's disc. A large portion of the Northern arm was already
discussed in \citet{Mackey2016}, where it was traced out to
$\sim20^{\circ}$ away from the LMC's center. Here we show that this
structure continues to higher $L_{\rm MS}$ for (at least) some
$5^{\circ}$ to $10^{\circ}$, passing right under the Carina dwarf,
where, as pointed in \citet{Mackey2016}, the LMC's stars have been
detected before \citep[see][]{Majewski2000, McMonigal2014}. In the
Southern regions of the LMC, a more complicated web of sub-structures
can be seen. There are two ``claw''-like over-densities, identified as
``Substructure 1'' and ``Substructure 2'' in
\citet{Mackey2018}. Curiously, in the maps presented here,
``Substructure 2'' appears to be curving clockwise, continuing as far
as $(L_{\rm MS}, B_{\rm MS})=(10^{\circ},-5^{\circ})$. One of the most
striking new features is a thin stellar stream which appears to be
connecting to the SMC at around $L_{\rm MS}\sim-8^{\circ}$. This
narrow tail, one of the longest structures discussed here, wraps
around the Southern edge of the LMC's disc, tracing an arc of
$\sim$90$^{\circ}$ in clockwise direction. As gleaned from
Figure~\ref{fig:map}, the LMC appears to have two long arms, one in
the North and its counter-part in the South.

\begin{figure}
  \centering
  \includegraphics[width=0.47\textwidth]{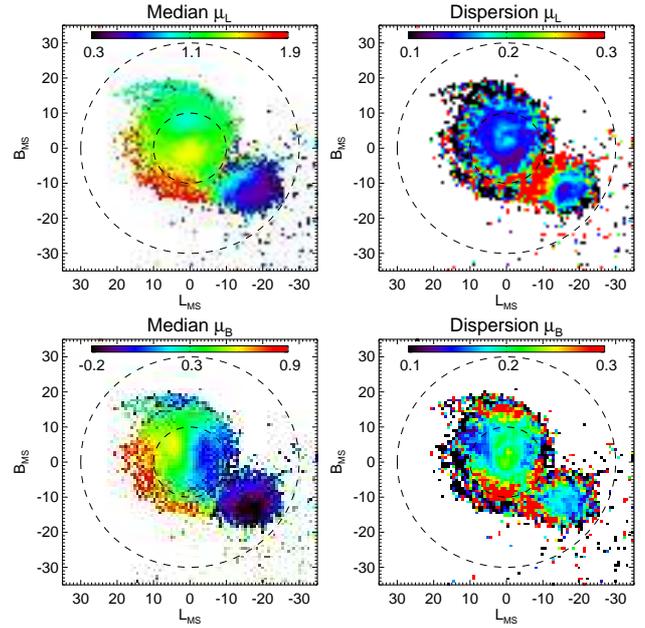}
  \vskip -0.2cm
  \caption[]{Kinematics of the selected RG stars. {\it Left Column:}
    Median values of $\mu_{\rm L}$ (top) and $\mu_{\rm B}$ (bottom) PM
    components in each pixel of $L_{\rm MS}, B_{\rm MS}$. {\it Right
      Column:} PM dispersion (around the median) maps corresponding to
    the median motion maps shown in the left column.}
   \label{fig:vel}
\end{figure}

To clarify the origin of the stellar over-densities described above,
Figure~\ref{fig:vel} gives the proper motions of the selected LMC's
candidate RGB stars. Note that these proper motions have been
corrected for the Solar reflex assuming a constant heliocentric
distance of 49.9 kpc. The pattern of the median proper motion values
(left column of the Figure) across the inner $10^{\circ}$ (smaller
dashed circle) is dominated by the gradient associated with the
Cloud's rotation \citep[see also][]{Vasiliev2018}. Note, however, that
the stellar motions preserve coherence well outside the central
LMC. More fascinating still, all of the narrow arm-like features at
distances beyond $\sim15^{\circ}$ also display coherent systematic
motions. Overall, the kinematics of the Northern and Southern arms
resembles that of the outer LMC's disc but off-set in orbital
phase. Note that the bulk of the Southern sub-structure shares the
proper motion of the LMC. This is especially evident in the lower left
panel, where the stellar streams have colors from green to red,
similar the LMC's disc, while the SMC is dark blue. While not the main
focus of this Letter, it is worth commenting briefly on the proper
motion pattern of the SMC. According to Figure~\ref{fig:vel}, the
SMC's systemic motion is in the direction away from the LMC,
i.e. towards negative $L_{\rm MS}$ and negative $B_{\rm MS}$,
consistent with previous measurements \citep[see
  e.g.][]{kallivayalil_lmc_pm}. Also visible are clear proper motion
gradients, whose direction is roughly aligned with the line connecting
the centers of the two Clouds. While this gradient could be modelled
as a rotation signal \citep[see e.g.][]{Helmi2018}, we suggest it
could instead be be interpreted as the evidence for the strong tidal
stretching of the SMC by the LMC \citep[see also][]{Zivick2018}.

\begin{figure*}
  \centering
  \vskip -0.2cm
  \includegraphics[width=\textwidth]{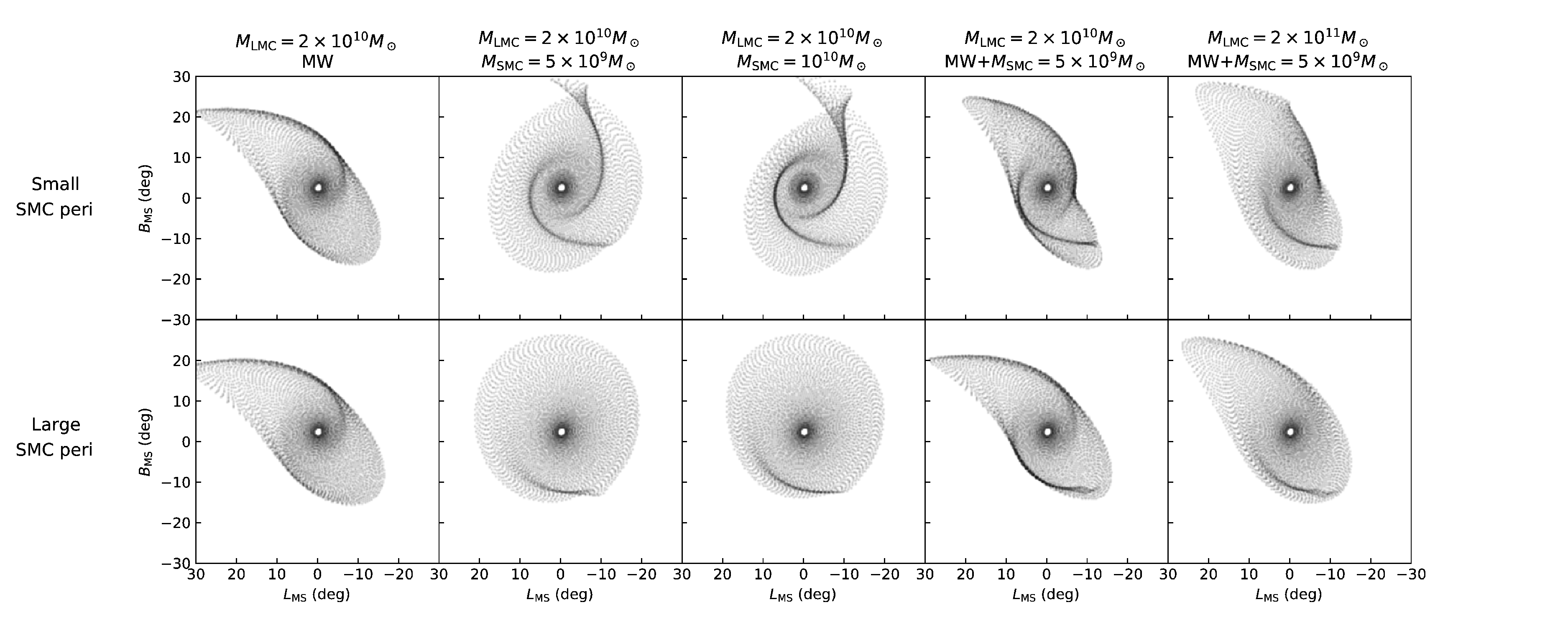}
  \vskip -0.2cm
  \caption[]{Simulated LMC disc in the presence of the MW and the SMC
    in Magellanic Stream coordinates. The left-most set of columns
    show the LMC evolved in the presence of only the MW. The
    second (third) set of columns shows the LMC evolved in the
    presence of only a $5\times 10^{9} M_\odot$ ($10^{11} M_\odot$)
    SMC. The fourth column shows the fiducial LMC evolved in the
    combined presence of the MW and a $5\times 10^9 M_\odot$
    SMC while the fifth column shows a $2\times 10^{11} M_\odot$ LMC
    evolved in the same setup. The two rows show two different
    realizations of LMC and SMC. The top (bottom) row shows an LMC
    with a closer (more distant) encounter with the SMC. These panels
    show that the MW and SMC are responsible for different
    features in the LMC disc. The tidal field of the MW
    predominantly bends the Northern part of the LMC disc, creating
    one of the spiral features. The influence of the SMC can create
    one or two spiral arms with the Southern spiral being more
    prominent. Together, the MW and SMC create two spirals as
    seen in the Gaia data (i.e. Fig. \protect\ref{fig:map}). Finally,
    the rightmost column shows that increasing the LMC mass results in
    a smaller deflection of the LMC disc due to the MW.}
   \label{fig:sim_multi}
\end{figure*}

The right column of the Figure presents dispersions around the median
values of proper motion components $\mu_{\rm L}$ and $\mu_{\rm B}$ for
each pixel of $L_{\rm MS}$ and $B_{\rm MS}$. Strong perturbations of
the inner LMC's disc have recently been reported in the literature
\citep[see][]{Choi2018}, but here, we offer a much more complete map
of kinematically cold (blue) and hot (red) regions across the entire
Cloud. The regions of elevated dispersion are clearly different for
the longitudinal and latitudinal proper motion components. For
$\mu_{\rm L}$, the hottest region is on the rim of the LMC's disc
facing the SMC and in between the Clouds, where one naturally expects
a mixture of stars from both dwarfs. In $\mu_{\rm B}$, there are two
extended regions with high proper motion dispersion, one in the North
and one in the South, located radially inward from the locations of
each arm. The arms themselves are distinctly cold as judged by their
dark blue color. 

\section{Simulations, Caveats and Conclusions}

In order to investigate how the MW and SMC affect the LMC's disc and
whether they can induce the spiral features shown in Figure
\ref{fig:map}, we have run a series of simulations in the spirit of
\cite{Toomre1972}. In particular, we simulate the disc of the LMC as a
series of particles in concentric rings which are initially on
circular orbits. The system is then evolved in the combined presence
of the MW and the SMC. We model the LMC's potential as a Hernquist
profile \citep{hernquist_profile} which satisfies the rotation curve
measurement at a radius of 8.7 kpc from \cite{vandermarel_lmc} (for
each LMC mass, the scale radius is fixed by this constraint). The
initial orientation and rotation sense of the LMC are chosen to match
the observations from \cite{vandermarel_lmc}. The SMC is also modelled
as a Hernquist profile which satisfies the rotation curve measurement
at a radius of 3 kpc from \cite{staminirovic_smc_mass}. The MW is
modelled as the 3-component potential, \texttt{MWPotential2014}, from
\cite{galpy}. Starting from their present day positions, the LMC and
SMC are rewound for 1 Gyr (in the combined presence of each other and
the MW), at which point particles are placed on circular orbits around
the LMC. For each simulation, we place 5000 particles on 50 separate
concentric circles (with radii evenly spaced from 1 to 20
kpc). The simulation is then evolved to the present. For the LMC's
present day position and velocity, we use a distance of $49.97 \pm
1.126$ kpc \citep{pietrzynski_lmc_dist}, a radial velocity of
$-262.2\pm3.4$ km/s \citep{vandermarel_lmc_rv}, and proper motions of
$(\mu_\alpha \cos \delta, \mu_\delta) = (1.91\pm 0.02,0.229\pm0.047)$
mas/yr \citep{kallivayalil_lmc_pm}. For the SMC's present day position
and velocity, we use a distance of $62.1\pm1.9$ kpc
\citep{graczyk_smc_distance}, a radial velocity of $145.6\pm0.6$ km/s
\citep{harris_smc_vel}, and proper motions of $(\mu_\alpha \cos
\delta, \mu_\delta) = (0.772\pm 0.063,-1.117\pm0.061)$ mas/yr
\citep{kallivayalil_lmc_pm}. For each choice of the LMC and SMC masses
and scale radii, we sample their present day position and velocity and
simulate 100 realizations to explore the variety of outcomes.

In Figure \ref{fig:sim_multi} we isolate the effect of the MW
(left-most column) and the SMC (middle two columns) on the LMC. The
two rows show two different realizations of the LMC and SMC's present
day position and velocity. The top row shows an LMC with a closer
encounter with the SMC ($r_{\rm peri} \sim 10$ kpc), while the bottom
row shows a more distant encounter ($r_{\rm peri} \sim 15$ kpc). The
tidal field of the MW is primarily responsible for bending the
Northern half of the LMC, similar to what was found in $N$-body
simulations in \cite{Mackey2016}, and creates a spiral arm feature
similar in position and orientation to what it seen in the data. The
SMC can create one or two spiral arms, depending on how strong of an
interaction it has with the LMC during its most recent
pericenter. While this is in seeming contradiction to the results of
\cite{Toomre1972}, we stress that we are observing the LMC disc only
$\sim 150$ Myr after its most recent passage with the SMC and that it
takes time for the spiral features to form. If the LMC was simulated
for longer, the second spiral would form in the lower panel of Figure
\ref{fig:sim_multi}. Interestingly, we find that the SMC creates a
strong spiral arm in the South which matches the observed Southern
stream. We found that changing the SMC mass from $5\times 10^9
M_\odot$ to $10^{10} M_\odot$ resulted in only a modest change in the
spiral features. Our simulations do not contain the ``claw''-like
features visible in the data in the Southern parts of the LMC. We
conjecture that these density features are remnants of much earlier
interactions between the two Clouds. The fourth column of the figure
shows the effect of both the MW and the SMC. This
demonstrates that their combined effect is needed to create the two
spirals observed in the LMC. It also illustrates how a close encounter
with the SMC can truncate the Western portion of the LMC's disc (top,
second from the right panel) similar to what it seen in the
data. Taken together, this shows that morphology of the LMC's disc and
the associated spiral structure can be used to reveal its rich
dynamical history.

In a similar vein, we also explore the effect of the LMC's mass on its
morphology in Figure \ref{fig:sim_multi}. While the first four columns
show a $2\times 10^{10}M_\odot$ LMC, the final column shows a $2\times
10^{11} M_\odot$ LMC.  Note that the final column should be compared
with the fourth column since these have the same setup.  As the LMC
mass is increased, we find that the LMC is deformed less by the
MW. This is because the increased LMC mass results in a larger tidal
radius and hence a larger region where the effect of the MW is
negligible. Interestingly, only the lowest mass LMC we consider
($2\times 10^{10} M_\odot$) can match the tightly wound spiral seen in
the North (see Fig. \ref{fig:map}). Since this mass is only slightly
higher than the mass constraint within 8.7 kpc from
\cite{vandermarel_lmc}, this could suggest that the LMC has already
been significantly stripped. However, we stress that these simulations
are only meant to be the first step in showing that the morphology
(including spirals) of the LMC's disc can provide useful constraints
on the properties of the LMC, SMC, and on the tidal effect of the
MW. With this aim in mind, the rich proper motions in Figure
\ref{fig:vel} will also be useful in future modelling efforts.

In summary, we have used the exquisite data from Gaia DR2 to
unveil a global view of the perturbations to the LMC's disc. In
particular, there are two clear spiral features, as well as some complex
substructure to the South of the LMC (see Fig. \ref{fig:map}). While
some of this structure was seen before
\citep[e.g.][]{Mackey2016,Mackey2018}, the uninterrupted view afforded
by Gaia allows us to better understand how these features
arose. We simulated the combined effect of the Milky Way and SMC on
the LMC's disc and found that both are important for creating the
spiral features seen in the data. Namely, the Milky Way is responsible
for deforming the Northern part of the LMC while the most recent
passage of the SMC creates the strong spiral feature in the South. A
close passage with the SMC can also truncate the Western side of the
LMC's disc. Finally, we showed that the distant Magellanic Red
Giants detected here can be used to map out the LMC's mass distribution at
unprecedentedly large distances.

\section*{Acknowledgments}

We thank Dougal Mackey, Adrian Price-Whelan, David Hogg, Gurtina
Besla, David Schiminovich, Alis Deason and Sergey Koposov for
illuminating discussions. The research leading to these results has
received funding from the European Research Council under the European
Union's Seventh Framework Programme (FP/2007-2013) / ERC Grant
Agreement n. 308024. This research was started at the NYC Gaia DR2
Workshop at the Center for Computational Astrophysics of the Flatiron
Institute in April 2018 and made use of data from the European Space
Agency mission Gaia (http://www.cosmos.esa.int/gaia), processed by the
Gaia Data Processing and Analysis Consortium (DPAC,
http://www.cosmos.esa.int/web/gaia/dpac/consortium). Funding for the
DPAC has been provided by national institutions, in particular the
institutions participating in the Gaia Multilateral Agreement.

\bibliography{references}

\label{lastpage}

\end{document}